\documentclass[sigconf]{acmart}
\pdfoutput=1

\usepackage{booktabs} 

\usepackage{multirow}
\usepackage{accents}
\usepackage{hyperref}
\usepackage{enumitem}

\setcopyright{rightsretained}

\copyrightyear{2018}
\acmYear{2018}
\acmConference[KDD 2018]{24th ACM SIGKDD Conference on Knowledge Discovery and Data Mining}{August 19--23, 2018}{London, UK}
\acmBooktitle{KDD 2018, August 19--23, 2018, London, UK}
\acmPrice{0.00}
\acmDOI{0}
\acmISBN{0}

\begin{document}
	\title{Online Evaluation for Effective Web Service Development}
	\renewcommand{\shorttitle}{Online Evaluation for Effective Web Service Development}

	\author{Roman Budylin}
	\affiliation{%
		\institution{Yandex; Moscow, Russia}
	}
	\email{budylin@yandex-team.ru}
	
	\author{Alexey Drutsa}
	\affiliation{%
		\institution{Yandex; Moscow, Russia}
	}
	\email{adrutsa@yandex.ru}
	
	\author{Gleb Gusev}
	\affiliation{%
		\institution{Yandex; Moscow, Russia}
	}
	\email{gleb57@yandex-team.ru}
	
	\author{Pavel Serdyukov}
	\affiliation{%
		\institution{Yandex; Moscow, Russia}
	    }
	\email{pavser@yandex-team.ru}

	\author{Igor Yashkov}
	\affiliation{%
		\institution{Yandex; Moscow, Russia}
	}
	\email{excel@yandex-team.ru}

	\begin{abstract}
Development of the majority of the leading web services and software products today is generally guided by data-driven decisions based on evaluation that ensures a steady stream of updates, both in terms of quality and quantity. Large internet companies use online evaluation on a day-to-day basis and at a large scale. The number of smaller companies using A/B testing in their development cycle is also growing. Web development across the board strongly depends on quality of experimentation platforms. In this tutorial, we overview state-of-the-art methods underlying everyday evaluation pipelines at some of the leading Internet companies.
Software engineers, designers, analysts, service or product managers --- beginners, advanced specialists, and researchers --- can  learn how to make web service development data-driven and do it effectively.
	\end{abstract}

%
%

%




\maketitle

\section*{Introduction}
\label{sec_Intro}
Nowadays, the development of most leading web services and software products, in general, is guided by data-driven decisions that are based on online evaluation which qualifies and quantifies the steady stream of web service updates. Online evaluation is widely used in modern Internet companies (like search engines~\cite{2014-WWW-Deng,2015-KDD-Hohnhold,2015-WWW-Drutsa}, social networks~\cite{2013-KDD-Bakshy,2016-KDD-Xu}, media providers~\cite{2013-KDD-Bakshy}, and online retailers) in permanent manner and on a large scale.
Yandex run more than 100 online evaluation experiments per day; Bing reported on more than 200 run A/B tests per day~\cite{2013-KDD-Kohavi}; and Google conducted more than 1000 experiments~\cite{2015-KDD-Hohnhold}.
The number of smaller companies that use A/B testing in the development cycle of their products grows as well. The development of such services strongly depends on the quality of the experimentation platforms. In this tutorial, we overview the state-of-the-art methods underlying the everyday evaluation pipelines.

At the beginning of this tutorial (which is a shorter version of~\cite{2018-WWWTut-Budylin}), we make an introduction to online evaluation and give basic knowledge from mathematical statistics (40 min, Section~\ref{sec_StatFound}).
Then, we share approaches for development of online metrics (50 min, Section~\ref{sec_OnlineMetrics}).
This is followed by 
rich industrial experiences on constructing of an experimentation pipeline and evaluation metrics: emphasizing best practices and common pitfalls (55 min, Section~\ref{sec_ExpPipeline}).
A large part of our tutorial is devoted to modern and state-of-the-art techniques (including the ones based on machine learning) that allow to conduct online experimentation efficiently (65 min, Section~\ref{sec_ML}). Finally, we point out open research questions and current challenges that should be interesting for research scientists.

\section{Statistical foundation}
\label{sec_StatFound}
We introduce the main probabilistic terms, which form a theoretical foundation of A/B testing. We introduce the observed values as random variables sampled from an unknown distribution. Evaluation metrics are statistics based on observations (mean, median, quantiles, etc.).
Overview of statistical hypothesis testing is provided with definitions of p-value, type I error, and type II error. We discuss several statistical tests (Student's t-test, Mann Whitney U, and Bootstrap~\cite{efron1994introduction}), compare their properties and applicability~\cite{2015-CIKM-Drutsa}.

\section{Development of online metrics}
\label{sec_OnlineMetrics}
We deeply discuss how to build evaluation metrics, what is the main ingredient in online experimentation pipeline.
First, we introduce the notion of an A/B test (also known as an online controlled experiment): it compares two variants of a service at a time, usually its current version (control) and a new one (treatment), by exposing them to two groups of users~\cite{2004-book-Peterson,2007-KDD-Kohavi,2009-DMKD-Kohavi}.
Main components of a metric are presented: key metric, evaluation statitistic, statistical significance test, Overall Evaluation Criterion (OEC)~\cite{2009-DMKD-Kohavi}, and Overall Acceptance Criterion (OAC)~\cite{2015-CIKM-Drutsa}.
The aim of controlled experiments is to detect the causal effect of service updates on its performance relying on an Overall Evaluation Criterion (OEC)~\cite{2009-DMKD-Kohavi}, a user behavior metric (e.g., clicks-per-user, sessions-per-user, etc.) that is assumed to correlate with the quality of the service.
We show that development of a new good metric is a challenging goal, since an appropriate OAC should possess two crucial qualities: \emph{directionality} (the sign of the detected treatment effect  should align with positive/negative impact of the treatment on user experience) and \emph{sensitivity} (the ability
to  detect the statistically significant difference when the treatment effect exists)~\cite{2012-KDD-Kohavi,2015-KDD-Nikolaev,2016-KDD-Poyarkov,2016-KDD-Deng}.
The former property allows to make correct conclusions on the system quality changes~\cite{2012-KDD-Kohavi,2015-KDD-Nikolaev,2016-KDD-Deng}, while improvement of the latter one allows to detect metric changes in more experiments and
 to utilize less users~\cite{2015-SIGIR-Kharitonov,2015-SIGIR-Kharitonov2,2016-KDD-Poyarkov}.

Second, we provide evaluation criteria beyond averages: (a) how to evaluate  periodicity~\cite{2017-TWEB-Drutsa,2015-SIGIR-Drutsa,2017-WWW-Drutsa} and trends~\cite{2015-WSDM-Drutsa,2015-SIGIR-Drutsa,2017-TWEB-Drutsa} of user behavior over days, e.g., for detection of delayed treatment effects; and (b) how to evaluate frequent/rare behavior and diversity in behavior between users that cannot be detected by mean values~\cite{2015-CIKM-Drutsa}.
Third, product-based aspects in metric building are presented.
Namely, we discuss vulnerability of metrics such as a click on a button to switch a search engine~\cite{2015-SIGIR-Arkhipova} (i.e., how can a metric be gamed or manipulated);
ways to measure different aspects of a service (i.e., speed~\cite{2010-MSWP-Kohavi,2014-KDD-Kohavi}, absence~\cite{2014-SIGIR-Chakraborty}, abandonment~\cite{2014-KDD-Kohavi});
difference between metrics of user loyalty and ones of user activity~\cite{2010-CHI-Rodden,lalmas2014measuring,2015-WSDM-Drutsa,2015-WWW-Drutsa,2015-SIGIR-Drutsa,2015-CIKM-Drutsa};
dwell time to improve click-based metrics~\cite{kim2014modeling};
how to evaluate the number of user tasks (which can have a complex hierarchy~\cite{boldi2009dango}) by means of sessions~\cite{2013-WWW-Song}; and issues in session division~\cite{jones2008beyond} as well.

Fourth, math-based approaches to improve metrics are considered. In particular, we discuss the powerful method of linearization~\cite{2018-WSDM-Budylin} that reduces any ratio metric to the average of a user-level metric preserving directionality and allowing usage of a wide range of sensitivity improvement techniques developed for user-level metrics.
We also describe different methods of noise reduction (such as capping~\cite{2014-KDD-Kohavi}, slicing~\cite{2013-WWW-Song,2015-WSDM-Deng}, taking into account user activity, etc.) and of utilization of a user generated content approach.

Finally, some system requirements for metric building are discussed. We explain how to get a set of experiments with verdicts (known positiveness or negativeness), how to construct a pipeline to easily implement and test metrics, and how to measure metrics~\cite{2016-CIKM-Dmitriev}.

\section{Experimentation pipeline and workflow in the light of industrial practice}
\label{sec_ExpPipeline}
We share rich industrial experiences on constructing of an experimentation pipeline in large Internet companies.
First, we discuss how can experiments be used for evaluation of changes in various components of  web services: the user interface~\cite{2009-IWDMCS-Kohavi,2015-WSDM-Drutsa,2015-KDD-Nikolaev,2017-TWEB-Drutsa}, ranking algorithms~\cite{2013-WWW-Song,2015-WSDM-Drutsa,2015-KDD-Nikolaev,2017-TWEB-Drutsa}, sponsored search~\cite{2016-EC-Chawla}, and mobile apps~\cite{2016-KDD-Xu}.
Second, we consider several real cases of experiments, where pitfalls~\cite{2009-KDD-Crook,2012-KDD-Kohavi,2014-KDD-Kohavi,2016-KDD-Deng} are demonstrated and  lessons  are learned. In particular, we discuss: conflicting experiments,  network effects~\cite{gui2015network}, duration and seasonality~\cite{shokouhi2011detecting,2015-WSDM-Drutsa}, logging, and slices.

Third, we provide our management methodology to conduct experiments efficiently and to avoid the pitfalls. This methodology is based on pre-launch checklists and a team of Experts on Experiments (EE). We also present our system of tournaments, where problems similar to the ones in two-stage A/B testing~\cite{2014-WWW-Deng} are solved.
Finally, we discuss how large-scale experimental infrastructure~\cite{2010-KDD-Tang,2013-KDD-Kohavi,2015-KDD-Xu} can be used to collect experiments for metric evaluation~\cite{2016-CIKM-Dmitriev}.

\section{Machine learning driven A/B testing}
\label{sec_ML}
A large part of our tutorial is devoted to modern and state-of-the-art techniques (including the ones based on machine learning) that improve the efficiency of online experiments.
We start this section with the comparison of randomized experiments and observational studies. We explain that the difference between averages of the key metric may be misleading when measured in an observational study. We introduce the Neyman--Rubin model and rigorously formulate implicit assumptions we make each time when evaluating the results of randomized experiments.

Then we overview several studies devoted to the variance reduction of evaluation metrics. Regression adjustment techniques such as stratification, linear models~\cite{2013-WSDM-Deng,2016-KDD-Xie}, and gradient boosted decision trees~\cite{2016-KDD-Poyarkov} reduce the variance related to the observed features (covariates) of the users.
We also consider experiments with user experience, where the effect of a service change is heterogeneous (is different for users of different types). We overview the main approaches to estimation of the heterogeneous treatment effect depending on the user features~\cite{powers2017some,athey2015machine}.

We explain the Optimal Distribution Decomposition (ODD) approach that is based on the analysis of the control and treatment distributions of the key metric as a whole,
and, for this reason, is sensitive to more ways the two distributions may actually differ~\cite{2015-KDD-Nikolaev}.
Method of virtually increasing of the experiment duration through the prediction of the future~\cite{2015-WWW-Drutsa} is discussed. We also provide another way to improve sensitivity that is based on learning of metric combinations~\cite{2017-WSDM-Kharitonov}. This approach showed outstanding sensitivity improvements in the large scale empirical evaluation~\cite{2017-WSDM-Kharitonov}.

Finally, we discuss ways to improve the performance of experimentation pipeline as a whole. Optimal scheduling of online evaluation experiments is presented~\cite{2015-SIGIR-Kharitonov2} and approaches for early stopping of them are highlighted (where inflation of Type I error~\cite{johari2017peeking} and ways to correctly make sequential testing~\cite{2015-SIGIR-Kharitonov,2016-DSAA-Deng} are discussed).

We also highlight important topics not covered by the tutorial:
Bayesian approaches~\cite{2015-WWWc-Deng,2016-DSAA-Deng} and non-parametric mSRPT~\cite{abhishek2017nonparametric} in sequential testing;
network A/B testing~\cite{gui2015network,saveski2017detecting};
two-stage A/B testing~\cite{2014-WWW-Deng};
Imperfect Treatment Assignment~\cite{coey2016people};
and interleaving~\cite{joachims2002unbiased,joachims2003evaluating,radlinski2008does,hofmann2011probabilistic,chapelle2012large,radlinski2013optimized,schuth2014multileaved,2015-CIKM-Kharitonov,netflix,Radlinski2011tutorial,Grotov2016,Radlinski2013,Radlinski2013SS}.


\section*{Tutorial materials}

The tutorial materials (slides) are available at  \url{https://research.yandex.com/tutorials/online-evaluation/kdd-2018}.


\bibliographystyle{ACM-Reference-Format}
\bibliography{2018-kdd-tutorial-extabstr}


\end{document}